\documentclass[twocolumn,showpacs,preprintnumbers,amsmath,amssymb,showkeys]{revtex4}

\usepackage{graphicx}
\usepackage{dcolumn}
\usepackage{bm}
\usepackage{bezier}

\begin{document}

\title{Dissipative work, Clausius inequality, and microscopic reversibility for open Hamiltonian systems}
\author{Takaaki Monnai}
\affiliation{Waseda Institute for Advanced Study, 1-6-1 Nishiwaseda Shinjuku-ku, Tokyo 169-8050, Japan}%

\date{\today}

\begin{abstract}
We derive a microscopic expression of the Clausius inequality for isothermal open systems within the fundamental Hamiltonian dynamics.
We also show the positivity of the dissipative work $\frac{{\cal W}_{diss}}{k_B}={\cal D}[P_F[\Gamma_s(t)]||P_R[\Theta\Gamma_s(T-t)]]$, which is a path integral generalization of the so-called relative entropy for the probability functionals of time forward- and reversed- trajectories. 
\end{abstract}
\pacs{05.30.-d,05.70.Ln}
\keywords{informational expression of the dissipative work, microscopic reversibility, open systems}
\maketitle
\section*{Introduction}
There are several expressions of the second law of thermodynamics\cite{Fermi,Vandenbroeck}.
In the context of the so-called perpetual motion of the second kind\cite{Vandenbroeck}, the Thomson's principle is useful. 
Recently, this principle has been experimentally tested for fluctuating systems such as a dragged colloidal particle in water\cite{Wang}, which was theoretically supported by Langevin stochastic analysis\cite{vanZon,Tasaki}.
These papers confirmed the positivity of the energy change by calculating the total work done. 
However, we also need the heat transfer and the internal entropy change to express the Clausius inequality.

Thus, it is desired to formulate 
the open system within the fundamental Hamiltonian dynamics, i.e., to trace out the experimentally inaccessible reservoir degrees of freedom.
Recently, the second law of thermodynamics have been pursued based on the Hamiltonian approaches\cite{Jarzynski1,Parrondo,Parrondo2,Monnai1}, which contain various other dynamics as special limiting cases\cite{Zwanzig}.
In particular, our previous paper\cite{Monnai1} contains Hamiltonian derivation of the so-called  microscopic reversibility for open systems\cite{Crooks1,Crooks2}, which connects the probability functional of the trajectory to the heat\footnote{The universality of the microscopic reversibility was confirmed by exploring the case of Nose-Hoover thermostatted dynamics\cite{Monnai2} with the use of a clear separation of the system and reservoir\cite{Esposito1}.}.
On the other hand, Kawai et al. \cite{Parrondo,Parrondo2} showed that the dissipation or entropy production is microscopically expressed for the {\it isolated systems}, which is in marked contrast to the  present article.   
Indeed, the Clausius inequality essentially provides an open system formulation of the second law of thermodynamics, which is expressed by the heat flowing into the system, the temperature of the reservoir, and the entropy change of the system between two different equilibrium states. 

In this article, we give a formal derivation of the Clausius inequality by considering the forward and corresponding reversed trajectories in phase space, which are sampled from initial canonical ensembles.  In particular, the derivation reveals a new informational expression of the dissipative work ${\cal W}_{diss}$ for {\it open} systems by a path integral generalization of the so-called relative entropy. The second law is, then, directly expressed as the positivity of this quantity, which is considered as an open system version of the dissipation obtained in Ref. \cite{Parrondo}.       
This paper is organized as follows.
   
\section*{Microscopic reversibility for open systems}
Let us briefly summarize the essence of the Hamiltonian derivation of the microscopic reversibility for open systems\cite{Crooks1,Crooks2,Crooks3} given in Eq. (30) of Ref. \cite{Monnai1}.
We consider an externally driven system, which weakly interacts with a thermodynamic reservoir at an inverse temperature $\beta$\footnote{For simplicity, we assume that the heat capacitance of the reservoir is large enough so that $\beta$ is regarded as constant}.
Then, we denote the set of the system- and reservoir- variables as $\Gamma_s$ and $\Gamma_r$, which consist of all the canonical coordinates and their conjugate momenta.              
Similarly, let us abbreviate the set of all the variables as $\Gamma=\{\Gamma_s,\Gamma_r\}$. 
The Hamiltonian of the total system $H(\Gamma,t)$ is decomposed to those of the system $H_s(\Gamma_s,t)$, reservoir $H_r(\Gamma_r)$ and interaction between them $V(\Gamma)$ as
\begin{equation}
H(\Gamma,t)=H_s(\Gamma_s,t)+H_r(\Gamma_r)+V(\Gamma).
\end{equation}
By the weak interaction, we mean that the interaction energy $V(\Gamma)$ is negligibly small compared with the bulk energies $H_s(\Gamma_s,t)$ and $H_r(\Gamma_r)$\footnote{
This condition would be weaker than the so-called weak coupling limit, where the absolute value of $V(\Gamma)$ is negligible.}.  
We assume that these Hamiltonians are invariant under the time reversal
$H(\Theta\Gamma,t)=H(\Gamma,t)$, $H_s(\Theta\Gamma_s,t)=H_s(\Gamma_s,t)$, and $H_r(\Theta\Gamma_r,t)=H_r(\Gamma_r,t)$, where $\Theta$ is the time reversal operator which reverses all the momenta\footnote{
In the presence of the magnetic field, we also reverse the direction of the magnetic field.}.
Since we apply an external forcing only to the system of interest, the system energy $H_s(\Gamma_s,t)$ is explicitly time dependent. 

First, we give definitions of the work and heat as a functional of the system state trajectory $\{\Gamma_s(t)\}$, which is consistent with those of Refs.\cite{Jarzynski1,Sekimoto}.
The work done during a time interval $0\leq t\leq T$ is the energy change due to the explicit time dependence of the system Hamiltonian $H_s(\Gamma_s(t),t)$ as
\begin{equation}
W[\Gamma_s(t)]\equiv\int_0^T\frac{\partial H_s(\Gamma_s(t),t)}{\partial t}dt.
\end{equation}
Then, the heat $Q[\Gamma_s(t)]$ flowing to the system from the reservoir is naturally defined by   the first law of thermodynamics. Note that the work done is just equal to the total energy change\cite{Jarzynski2} 
\begin{equation}
W[\Gamma_s(t)]=H(\Gamma(T),T)-H(\Gamma(0),0). \label{energychange}
\end{equation}
Indeed, by requiring the first law $W[\Gamma_s(t)]+Q[\Gamma_s(t)]=H_s(\Gamma_s(T),T)-H_s(\Gamma_s(0),0)$, Eq. (\ref{energychange}) just yields  
\begin{equation}
Q[\Gamma_s(t)]=-\left(H_r(\Gamma_r(T))-H_r(\Gamma_r(0))\right) \label{heat1}
\end{equation}
due to the weak coupling\footnote{One may add the coupling energy change $V(\Gamma(T))-V(\Gamma(0))$, however, this is negligible compared with Eq. \ref{heat1}.}.   
In short, the heat flowing to the system is the explicit energy change of the reservoir.
Experimentally, the heat was also measured for the mesoscopic systems\cite{Ciliberto}.
In order to discuss on the microscopic reversibility, we need the joint probability functionals $P_F[\Gamma_s(t)]$ and $P_R[\Theta\Gamma_s(T-t)]$ to have a specific trajectory of the system state $\{\Gamma_s(T)\}$ and its time reversal $\{\Theta\Gamma_s(T-t)\}$ for the forward and time reversed forcing protocols. For the precise of the definitions, see Ref. \cite{Monnai1} and Ref.  \cite{Monnai2}.
The interaction with the reservoir acts as a kind of thermal noise, and the system state trajectory fluctuates.
And we can consider the probability of each trajectory.   
Here we precisely define the forward and time reversed protocols.
\subsection{Forward protocol}
Initial state $\Gamma(0)$ is sampled from an ensemble $\rho(\Gamma,0)=\rho_s(\Gamma_s,0)\rho_r^{eq}(\Gamma_r)$, where $\rho_r^{eq}(\Gamma_r)\equiv\frac{1}{Z_r}e^{-\beta H_r(\Gamma_r)}$ is the canonical ensemble.
Here, the ensemble is given as the product due to the weak interaction.
$Z_r\equiv\int d\Gamma_r e^{-\beta H_r(\Gamma_r)}$ is the partition function of the reservoir.
For the derivation of the microscopic reversibility, $\rho_s(\Gamma_s,0)$ can be arbitrary.
(But for the derivations of the Clausius inequality and the informational expression of the dissipative work in Sec. 3 and 4, we assume that $\rho_s(\Gamma_s,0)$ is also canonical. In particular, for the formal derivation of the Clausius inequality (\ref{entropychange1}), we further require that both the initial and final states are described by the canonical ensembles. This additional assumption is inevitable, since there should be {\it equilibrium entropy change}.
Remarkably, however, the positivity of the dissipative work ${\cal W}_{diss}$ (\ref{entropy1}) is derived without this additional assumption on the thermalization.)   
 
\subsection{Reversed protocol}
We assume that the initial state of time reversed protocol is statistically independent from the forward trajectory\cite{Jarzynski1}, and only the external forcing protocol is reversed  with respect to time $T$.
Then, the time reversed protocol is similarly defined as in the forward protocol.
The initial state $\Theta\Gamma(T)$ is sampled from an ensemble
$\tilde{\rho}_s(\Theta\Gamma_s,T)\rho_r^{eq}(\Theta\Gamma_r)$.
Here, tilde emphasizes that the initial distributions $\rho_s(\Gamma_s,0)$ and $\tilde{\rho}_s(\Theta\Gamma_s,T)$ are mutually independent. 
As mentioned earlier, the reservoir Hamiltonian is time reversal symmetric, and $\rho_r^{eq}(\Theta\Gamma_r)=\frac{1}{Z_r}e^{-\beta H_r(\Gamma_r)}$.
We denote the time reversed trajectory as $\{\Theta\Gamma_s(T-t)\}$, since the momenta change their sign and the forcing protocol are reversed with respect to the time $T$. 

Let us define the conditional probability $P_F[\Gamma_s(t):\Gamma_s(0)]\equiv\frac{1}{\rho_s(\Gamma_s(0),0)}P_F[\Gamma_s(t)]$ to have a trajectory $\{\Gamma_s(t)\}$ provided that the initial system state was $\Gamma_s(0)$ and the corresponding conditional probability for the reversed process $P_R[\Theta\Gamma_s(T-t):\Theta\Gamma_s(T)]\equiv\frac{1}{\tilde{\rho}_s(\Theta\Gamma_s(T),T)}P_R[\Gamma_s(T-t)]$.
These conditional probabilities satisfy the microscopic reversibility principle\cite{Crooks1,Crooks2}
\begin{equation}
\frac{P_F[\Gamma_s(t):\Gamma_s(0)]}{P_R[\Theta\Gamma_s(T-t):\Theta\Gamma_s(T)]}=e^{-\beta Q[\Gamma_s(t)]}. \label{microscopic}
\end{equation}        
\section*{Clausius inequality}
Let us go to the main part of this article, i.e., a formal derivation of the Clausius inequality based on the microscopic reversibility for the Hamiltonian dynamics Eq. (\ref{microscopic}).
First, we note that Eq. (\ref{microscopic}) provides the microscopic expression of the heat
\begin{eqnarray}
-\beta Q[\Gamma_s(t)]&=&\log\frac{P_F[\Gamma_s(t):\Gamma_s(0)]}{P_R[\Theta\Gamma_s(T-t):\Theta\Gamma_s(T)]} \nonumber \\
&=&\log\frac{P_F[\Gamma_s(t)]}{P_R[\Theta\Gamma_s(T-t)]}\frac{\tilde{\rho}_s(\Theta\Gamma_s(T),T)}{\rho_s(\Gamma_s(0),0)}. \label{heat1} \nonumber \\
&&
\end{eqnarray}
Now, our central assumption is that the initial distribution functions of the system $\rho_s(\Gamma_s(0),0)$ and $\tilde{\rho}_s(\Theta\Gamma_s(T),T)$ for the forward and time reversed processes are described by the canonical ensembles at the same inverse temperature $\beta$\cite{Jarzynski1,Monnai1}. 
Since the Clausius inequality connects two different equilibrium states, we here additionally assume that the final state of the forward process is in equilibrium under the interaction with  the reservoir.
We want to stress that the thermalization assumption is not essential for the second law\cite{Parrondo2}.
Actually, it is not necessary to derive another version of the second law, i.e., the nonnegativity of the dissipative work as we will show later in Eq. (\ref{entropy1}). 
In short, we use the thermalization assumption only for the formal expression of the Clausius inequality. 

Then, the derivation is straightforward.     
From the time reversal invariance of the system Hamiltonian, we have  $\tilde{\rho}_s(\Theta\Gamma_s(T),T)=\tilde{\rho}_s(\Gamma_s(T),T)$.
By taking the sample average of the heat Eq.(\ref{heat1}) and using the initial canonical conditions, we have the Clausius inequality 
\begin{eqnarray}
&&\langle-\beta Q[\Gamma_s(t)]\rangle \nonumber \\
&=&\int D\Gamma_s(t)P_F[\Gamma_s(t)]\log\frac{P_F[\Gamma_s(t)]}{P_R[\Theta\Gamma_s(T-t)]}\frac{\tilde{\rho}_s(\Theta\Gamma_s(T),T)}{\rho_s(\Gamma_s(0),0)} \nonumber \\
&=&\int D\Gamma_s(t)P_F[\Gamma_s(t)]\bigl(-\beta(H_s(\Gamma_s(T),T)-H_s(\Gamma_s(0),0)  \nonumber \\
&&-\Delta F_s)+\log\frac{P_F[\Gamma_s(t)]}{P_R[\Theta\Gamma_s(T-t)]}\bigr) \nonumber \\
&=&-\beta\left(\langle\Delta H_s(\Gamma_s(u))|_{u=0,T}\rangle-\bigl(\Delta E-\frac{\Delta S}{k_B\beta}\bigr)\right) \nonumber \\
&&+\int D\Gamma_s(t)P_F[\Gamma_s(t)]\log\frac{P_F[\Gamma_s(t)]}{P_R[\Theta\Gamma_s(T-t)]} \nonumber \\
&=&-\frac{\Delta S}{k_B}+{\cal D}[P_F[\Gamma_s(t)]||P_R[\Theta\Gamma_s(T-t)]] \nonumber \\
&\geq & -\frac{\Delta S}{k_B}. \label{entropychange1}
\end{eqnarray}
In the average $\langle A[\Gamma_s(t)]\rangle\equiv\int D\Gamma_s(t)P_F[\Gamma_s(t)]A[\Gamma_s(t)]$, the trajectories connect initial and final canonical distributions, and statistically fluctuate due to the interaction to the reservoir.
Here $k_B$ is the Boltzmann constant.
In the first equality, the sample average is expressed by the path integral.
In the second equality, we used the assumption of initial canonical states.
We also decompose the equilibrium free energy change to the energy change and contribution from the entropy change $\Delta F=\Delta E-\frac{\Delta S}{k_B\beta}$, where the energy change $\Delta E$ just cancels the averaged energy change $\langle\Delta H_s(\Gamma_s(u))|_{u=0,T}\rangle=\langle H_s(\Gamma_s(T))-H_s(\Gamma_s(0))\rangle$.
The inequality comes from the non-negativity of 
\begin{eqnarray}
&&{\cal D}[P_F[\Gamma_s(t)]||P_R[\Theta\Gamma_s(T-t)]] \nonumber \\
&\equiv& \int D\Gamma_s(t)P_F[\Gamma_s(t)]\log\frac{P_F[\Gamma_s(t)]}{P_R[\Theta\Gamma_s(T-t)]},
\end{eqnarray} 
which is a path integral generalization of the relative entropy.
    
The equality holds for the quasi static limit, where the forward and reversed probabilities are trivially equal $P_F[\Gamma_s(t)]=P_R[\Theta\Gamma_s(T-t)]$.
Eq.(\ref{entropychange1}) actually implies the second law for the mesoscopic and macroscopic systems
, since $\Delta S$ is the entropy change of the system, and $\Delta S_r\equiv \langle -\beta Q[\Gamma_s(t)]\rangle$ is that of the reservoir.
For the total system, the entropy is increasing $\Delta S+\Delta S_r\geq 0$ in accordance with the second law of thermodynamics.
\section*{Dissipation}
As mentioned earlier, here we don't require that the final state is in equilibrium.
Let us consider the dissipative work $\beta {\cal W}_{diss}\equiv\beta(\langle W[\Gamma_s(t)]\rangle-\Delta F_s)$, which is equal to the entropy production divided by $k_B$ when both the initial and final states are in equilibrium.  This is easily verified with the use of  $\Delta F_s=\Delta E-\frac{\Delta S}{k_B\beta}$, the first law $\langle W[\Gamma_s(t)]\rangle -\Delta E=-\langle Q[\Gamma_s(t)]\rangle$, and the definition of the entropy production $\Delta_i S=-\frac{\langle Q[\Gamma_s(t)]\rangle}{k_B}+\Delta S$. 

In general, the dissipative work measures an amount of the dissipation.
We show that the dissipative work is always nonnegative, which is another expression of the second law.
The derivation is based on the Eq. (\ref{entropychange1}).
In the third equality of Eq. (\ref{entropychange1}), we replace the energy change of the system by the sum of the work and heat $\langle \Delta H_s(u)|_{u=0,T}\rangle=\langle W[\Gamma_s(t)]\rangle +\langle Q[\Gamma_s(t)]\rangle$. Then, it is straightforward to show that the dissipative work satisfies
\begin{equation}
\beta{\cal W}_{diss}={\cal D}[P_F[\Gamma_s(t)]||P_R[\Theta\Gamma_s(T-t)]].\label{entropy1}
\end{equation}
Eqs. (\ref{entropychange1}) and (\ref{entropy1}) are the main results of this article.

In Ref. \cite{Parrondo,Parrondo2} the dissipation $\langle W[\Gamma_s(t)]\rangle-\Delta F_{total}$ is considered, where $\Delta F_{total}$ is the free energy change of the total system.
On the other hand, we are interested in $\langle W[\Gamma_s(t)]\rangle-\Delta F_{total}$.
But this difference would disappear in the weak interaction regime, since the free energy change of the reservoir should be negligible in such cases.
Also, an analogy to the action functional and entropy production for the Markovian stochastic processes in Ref. \cite{Lebowitz3} is clear due to the path integral formulation.  
In this context, it is remarked that the ratio between $P_F[\Gamma_s(t)]$ and $P_R[\Theta\Gamma_s(T-t)]$ is directly connected to the dissipative work for each trajectory $\{\Gamma_s(t)\}$
\begin{equation}
\frac{P_F[\Gamma_s(t)]}{P_R[\Theta\Gamma_s(t)]}=e^{\beta(W[\Gamma_s(t)]-\Delta F_s)}.\label{dissipation1}
\end{equation} 
In this way, expressions of the second law Eqs. (\ref{entropychange1},\ref{entropy1},\ref{dissipation1}) are obtained in terms of the underlying microscopic reversibility principle for open systems.
  
\section*{Discussion}
We derived two mutually related but different expressions of the second law for open Hamiltonian dynamics based on the microscopic reversibility. 
First, we derived an expression of the Clausius inequality.
The difference between the heat divided by the temperature $-\langle\beta Q[\Gamma_s(t)]\rangle$ and the entropy change $\frac{\Delta S}{k_B}$ is the entropy production in accordance with the second law.
Then, we showed the non-negativity of the dissipative work, which is expressed by a functional generalization of the relative entropy.
Microscopic reversibility for Hamiltonian systems\cite{Monnai1} also provides a path dependent expression of the dissipative work Eq. (\ref{dissipation1}).

These expressions are essentially an open system generalization for those of Kawai et al.\cite{Parrondo}.
It is interesting that, depending on whether the system is thermally isolated or open, the dissipation is expressed either by the relative entropy or its functional generalization.
Also, we quantify the dissipation as $\beta(\langle W\rangle-\Delta F_s)$, however, Refs. \cite{Parrondo,Parrondo2} are concerned $\beta(\langle W\rangle-\Delta F_{total})$.
The difference between these quantities is $\Delta F_r\equiv\Delta F_{total}-\Delta F_s$.
One might expect, that $\Delta F_r$ is small compared with $\Delta F_{total}$ for the weak interaction regime. But this would be model dependent, and further investigation of this issue is a future task.           
As in the case of the relative entropy\cite{Esposito1}, the functional generalization measures a closeness between the two different probability functionals. In our case, we are interested in the "distance" between the probability functionals $P_F[\Gamma_s(t)]$ and $P_R[\Theta\Gamma_s(T-t)]$ of the forward and reversed trajectories.  
\section*{Acknowledgment}
The author is grateful to Prof.Sugita for fruitful discussions.
This work is supported by Grants for special projects from Waseda University and JSPS Research program under the grant $22\cdot 7744$.

\end{document}